\documentclass[twocolumn,secnumarabic,amssymb, nobibnotes, prb]{revtex4-1}

\usepackage{color}
\usepackage{graphicx}
\usepackage{subfigure}
\usepackage{multirow}

\begin{document}
\pagestyle{myheadings}
\markright{Gr{\"u}nebohm {\it{et al.}} Ferroelectric trends of TiO$_2$} 
\author{A. Gr\"unebohm}
\email[Electronic mail: ]{anna@thp.uni-duisburg.de}
\affiliation{Faculty of Physics and Center for Nanointegration, CeNIDE, University of Duisburg-Essen, 47048 Duisburg, Germany}
\author{M. Siewert}
\affiliation{Faculty of Physics and Center for Nanointegration, CeNIDE, University of Duisburg-Essen, 47048 Duisburg, Germany}
\author{C. Ederer}
\affiliation{School of Physics, Trinity College, Dublin 2, Ireland}
\author{P. Entel}
\affiliation{Faculty of Physics and Center for Nanointegration, CeNIDE, University of Duisburg-Essen, 47048 Duisburg, Germany}
\title{First-principles study of the influence of (110) strain on the  ferroelectric trends of TiO$_2$}%
\begin{abstract}
We investigate the impact of uniaxial strain on atomic shifts,
dipolar interactions, polarization and electric permittivity
in TiO$_2$ (rutile) by using two different implementations of 
density functional theory. It is shown that calculations using the 
Vienna ab inito simulation package (VASP) and the plane-wave self-consistent field
method (PWscf) yield qualitatively the same atomic relaxations and ferroelectric
trends under strain. The phonon dispersion curves of unstrained
and strained TiO$_2$ (rutile) obtained by employing the linear response
method confirm previous calculations of the giant LO-TO splitting
and the appearance of soft polar modes. A second order phase transition into a ferroelectric phase with polarization along (110) appears under expansive strain in (110) direction. 
\end{abstract}
\maketitle
\section{Introduction}
TiO$_2$ is one of the key materials within oxidic systems.
Due to its large bio-compatibility and the well optimized and cheap processing, it is used in a huge variety of applications, e.g., see Tab.~\ref{tab:anwendungen}.
The first class of applications is based on its high photocatalytic power under UV-light.\cite{Graetzel, Clean,Air}\\
Additionally, TiO$_2$ is transparent for visible light and possesses a  large permittivity and therefore high refraction indices and is thus a very powerful material for optical tasks.\cite{Siefering}
The large permittivity also gives rise to applications of TiO$_2$ in microelectronics, for example as thin film capacitors. \cite{Wu}
Especially the  rutile morphology of TiO$_2$ is important which is an {\it{incipient}} ferroelectric material.\cite{Lee}
Although, no ferroelectric transition has been observed experimentally until now, its dielectric constant in (001) direction is large and increases drastically with reduced temperatures \cite{Samara} which indicates that the system is very close to a ferroelectric transition. 
The anomalous behavior of the dielectric constant can be linked to the softening of the optical A$_{2u}$ mode, which is the only polar mode in (001) direction. \cite{Lee} 
Theoretical investigations have shown further softening of this mode under lattice expansion or strain.
\cite{Montanari,Montanari2,Liu,felich,110,Mitev}
TiO$_2$ may be even interesting in the context of multiferroics since magnetic properties can be induced by doping with magnetic ions\cite{Co-DMS} or when  nanocomposite systems are formed. For example, magneto-electric coupling at Fe-TiO$_2$ interfaces has been observed experimentally.\cite{Yoon}
This would lead to new multifunctional devices, see Tab.~\ref{tab:anwendungen}.

Most existing applications are based on rutile surfaces and interfaces, especially on the rutile (110) surface.
This surface can be  strained in ($\bar{1}$10) or (001) direction by the growth on a proper substrate.
While a detailed discussion of the influence of (001) strain and isotropic expansion exists, the influence of (110) strain is still not completely understood.
For example, the impact of (110) strain  on the A$_{2u}$ mode is still under debate as this mode has been predicted to be insensitive with respect to such strain in Ref.~\onlinecite{Mitev}, whereas we recently found a softening of the mode.\cite{110}
Besides the destabilization of a "polar mode" in (001) direction corresponding to a strong ferroelectric tendency, other polar phonon modes in  (110)/($\bar{1}10$) direction soften due to expansive (001) or (110) strain.\cite{Liu,110}
We found out that a stable ferroelectric state with a large polarization in (110) direction can be stabilized under expansive strain along (110) within our {\it{ab initio}} simulations.\cite{110}
Hence, stable polar states can be expected in both, (001) and (110) direction, which show different dependencies on the magnitude of the imposed strain and its direction.
Also, a coupling between these soft modes exists.\cite{110}
Because of this, the possibility of "strain engineering"  the magnitude and direction of the polarization or at least  of the dielectric and the piezoelectric responses, due to different relative magnitudes of imposed (001) and (110) strain, is very likely.\\
Nevertheless, a detailed understanding of the strong polar tendency in (110) direction is still missing and our previous results are in part contradicting the theoretical results of Ref.~\onlinecite{Mitev}, where the destabilization of the structure along an acoustic phonon mode, or a non-polar phonon mode has been predicted under (110) strain.\\
The main goal of the present paper is to gain understanding of the influence of (110) strain on the phonon spectrum and especially on the polar modes polarized in (110) direction.
For this purpose, {\it{ab initio}} simulations have been performed with two different pseudopotential codes in order to cross-check our results.
The atomic displacements and the ferroelectric properties have been monitored for different values of expansive (110) strain.
Within both kind of simulations we find a ferroelectric instability of the system along the lowest polar mode in (110) direction. The phase transition to a ferroelectric state with large polarization is of second order and its onset is below 0.5~\% strain. Additionally, the lowest polar mode in (001) direction also softens, but to a lesser extent.

\begin{table*}[t]
\caption{Devices based on TiO$_2$ and future perspectives.}
\begin{tabular}{llll}
\hline
\hline
Photocatalytic properties& Optical properties&DMS \footnote{Co doped TiO$_2$ works as transparant diluted magnetic semiconductor (DMS) in Ref.~\onlinecite{Co-DMS}}\footnote{A review on oxidic DMS materials can be found in Ref.~\onlinecite{Fukumura}}/Electronic devices&ME\footnote{See review \onlinecite{Fiebig} on magneto-electrical coupling (ME) and possible devices.} and multiferroic properties\\
\hline
Solar cell\cite{Graetzel}&Anti-reflection coating &Magnetooptic devices  & Modulation of optical waves\\
Self-cleaning material \cite{Clean}&Pigments&MRAM\footnote{Magnetic Random Access Memory}& ME switching\\
Cancer therapy\cite{Cancer}&Interference filters \cite{Siefering}&Spintransistor&ME data storage\\
Air purification\cite{Air}&Optical waveguides \cite{Siefering}&$\dots$&Amplification\\
Sensors\cite{Gas}&Sun protection&Thin film capacitors \cite{Wu}&Frequency conversion\\
\hline
\hline
\end{tabular}
\label{tab:anwendungen}
\end{table*}
\section{Computational details}

All results have been obtained by self-consistent density functional theory calculations employing the plane wave pseudopotential codes VASP (Vienna Ab Initio Simulation Package) \cite{Kresse1} and the PWscf (Plane-Wave Self-Consistent Field) code which is part of the quantum espresso package.\cite{espresso}
Within the VASP simulations the projector augmented wave potentials \cite{Blochl} with the generalized gradient approximation (GGA) in the formulation of Perdew, Burke and Ernzerhof \cite{PBE} for the exchange correlation potential have been used. For Ti (O) atoms $4s3d$ ($2s 2p$) electrons have been treated as valence in the calculations. 
Within the PWscf simulations ultrasoft pseudopotentials with the local density approach (LDA) and Ti (O) $4s 3d 3s 3p$ ($2s 2p$) valence electrons have been used.
All ionic positions have been relaxed until forces converged to 0.001 eV/{\AA}.
Born charges, the dielectric permittivity and phonon frequencies have been calculated using density functional perturbation theory (DFPT).\cite{bornstoer} 
The "modern" theory of polarization \cite{Berry} has been used within our VASP simulation in order to determine the spontaneous polarization.
For the calculation of the phonon spectrum a Monkhorst-Pack\cite{Monkhorst} k-point grid containing 126 points was used leading to a convergency of at least 0.1~cm$^{-1}$ for selected special points of the Brillouin zone.
The phonons were calculated using $4\times4\times4$ $q$-point grid.

\section{Results}
\begin{figure*}
\subfigure[]{
\includegraphics[width=0.4\textwidth]{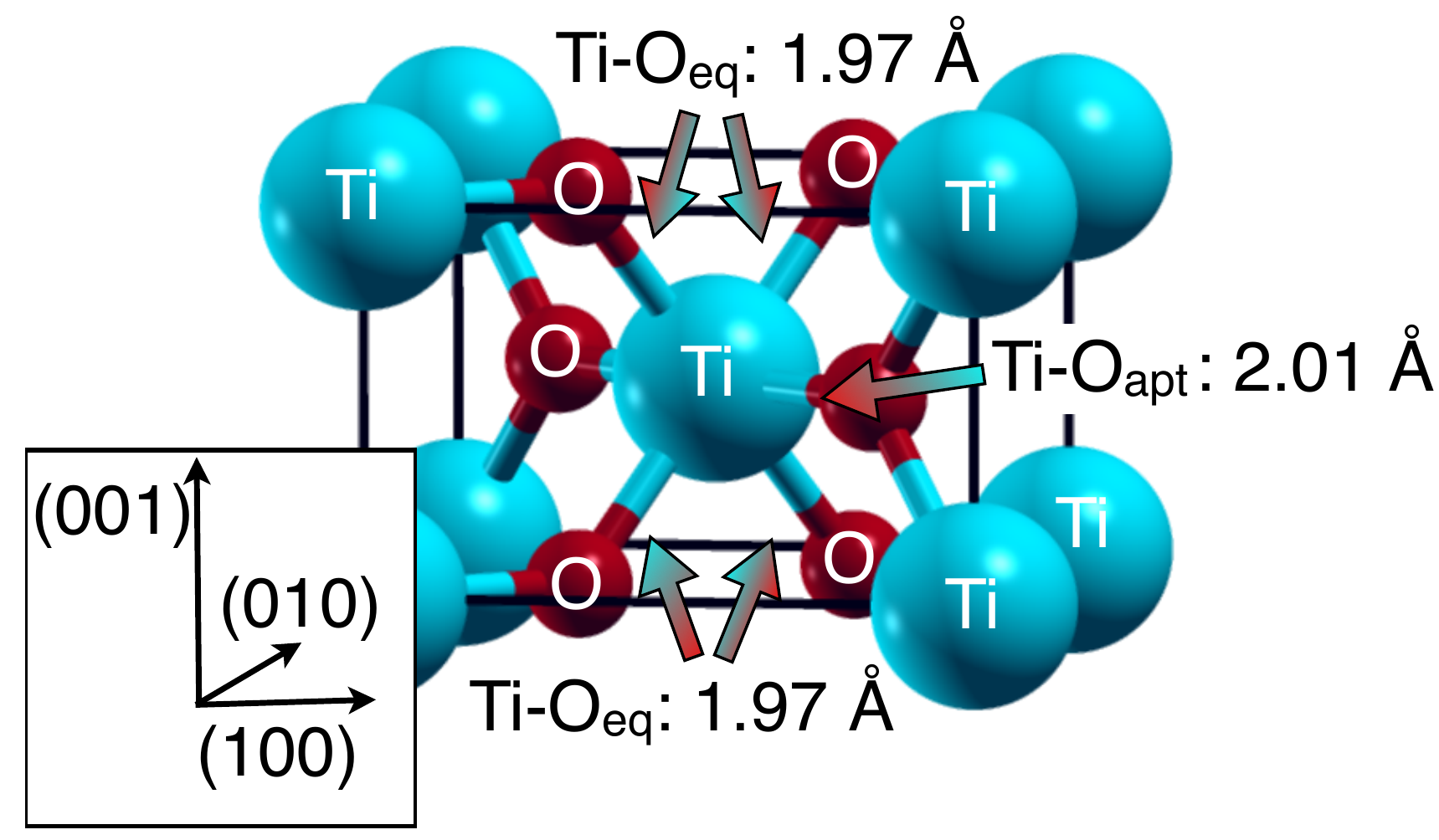} }
\subfigure[]{\includegraphics[width=0.3\textwidth]{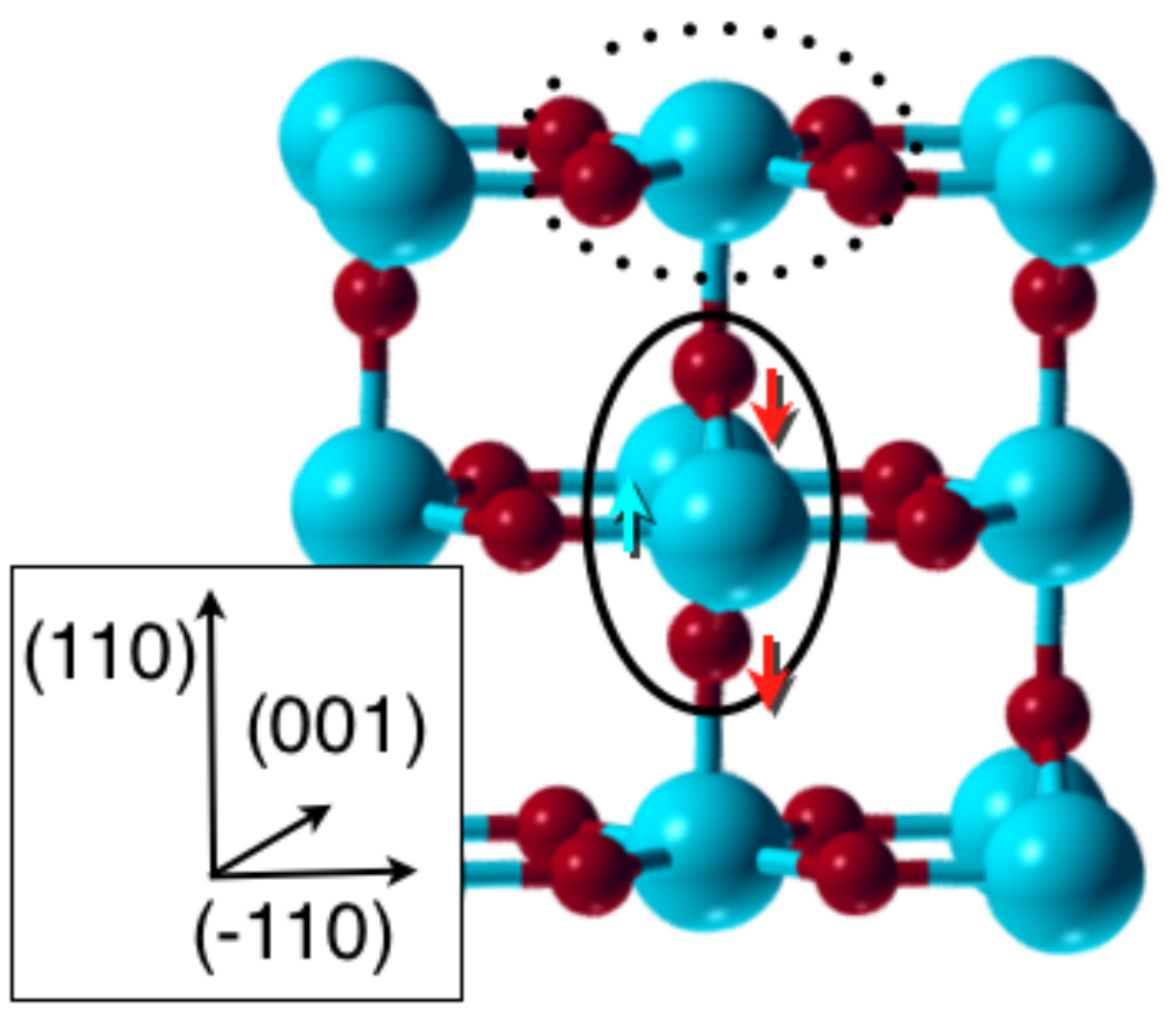} }
\caption{Atomic structure of TiO$_2$ rutile. The approximate atomic size is fixed by covalent radii. Ti: light, cyan, O: dark, red.
 (a) Primitive unit cell. (b) $\sqrt{2}\times\sqrt{2}\times1$ supercell. Ovals mark the different sublattices under (110) strain. Dotted line: class 1; solid line: class 2, see text. Arrows represent the main atomic shifts under (110) strain for one subclass.
}
\label{fig:rutil}
\end{figure*}
The rutile morphology of TiO$_2$ possesses $P4_2/mnm$ symmetry. 
Each Ti-atom is 6-fold coordinated by O-atoms with 4 short {\it{equatorial}} Ti-O bonds and two longer {\it{apical}}  bonds, see Fig.~\ref{fig:rutil}. 
Besides the lattice constant $a$ and the tetragonal ratio $c/a$, an internal parameter $u$ defines the relative length of both kinds of Ti-O distances.
The values, which have been obtained for these lattice constants agree with former results from literature, see Tab.~\ref{tab:para}. While the LDA lattice volume is slightly underestimated in comparison to the experimental values, it is overestimated in GGA calculations.
Although, this enlarged atomic volume may also lead to an overestimation of the ferroelectric character, the quantitative trends can be properly described by using GGA potentials.\cite{felich,Montanari2,Shojaee}
\begin{table}
\caption{Lattice parameters of bulk rutile. $a$: lattice constant; $c/a$: tetragonal ratio; $u$: internal parameter; $Z^*_i$ corresponds to the principal values of the Born tensor for Ti, along (001), along the apical Ti-O bond ($||$) and perpendicular to these directions. $\epsilon_{(001)}^{\infty}$ and $\epsilon_{(p)}^{\infty}$ is the permittivity along (001) respectively in the (010)/(100) plane.}
\begin{tabular}{cccccc}
\hline
\hline
&VASP&PWscf&Ref.~\cite{Lee}&Ref.~\cite{Mitev}&Exp.\\
\hline
a [{\AA}]&4.664&4.581&4.536&4.572&4.594\footnote{Ref.~\cite{Abrahams}}\\
c/a&0.637&0.641&0.643&0.644&0.644\footnotemark[1]\\
u&0.305&0.305&0.304&0.304&0.305\footnotemark[1]\\
$Z^*_{(001)}$&8.15&7.69&7.54&7.77&-\\
$Z^*_{||}$&7.60&7.26&7.34&7.49&-\\
$Z^*_{\perp}$&5.36&5.26&5.43&5.43&-\\
$\epsilon_{(001)}^{\infty}$&9.78&8.95&8.67&9.21&8.43\footnote{Ref.~\cite{Traylor}}\\\
$\epsilon_{(p)}^{\infty}$&7.95&7.46&7.54&7.95&6.84\footnotemark[2]\\
\hline
\hline
\end{tabular}

\label{tab:para}
\end{table}

\begin{table*}
\caption{Upper part: Frequencies of the optical phonon modes at $\Gamma$ for undistorted rutile and the deviations between our PWscf simulations and experimental results.
Lower part: Frequencies of the polar modes under 2~\% expansive strain along (110).}
\begin{tabular}{ccccccccc}
\hline
\hline
&A$_{2u}$ (TO)& A$_{2u}$(LO) &E$_{u,1}$ (TO)&E$_{u,2}$ (TO)&E$_{u,3}$ (TO)&E$_{u,1'}$ (LO)&E$_{u,2'}$ (LO)&E$_{u,3'}$ (LO)\\
\hline
Ref.\cite{Lee}&176.1&769.3&164.8&391.3&492.8&351.5&441.7&808.4\\
Exp.&144*&811&183&388&500&375&429&842\\
Ref.\cite{Mitev}&126&749&127&375&476&342&432&782\\
Ref.\cite{Montanari2}&47&-& 127&357&472&-&-&-\\
PWscf&38&743&90&367&475&355&429&765\\
Deviation[\%]&-73&-8&-50&-5&-5&-4&0&-6\\
\hline
\multirow{3}{*}{2~\% strain}&B$_{1u}$ (TO) &B$_{1u}$ (LO) &B$_{2u,1}$(TO)&B$_{2u,2}$(TO)&B$_{2u,3}$(TO)&B$_{2u,1'}$(LO)&B$_{2u,2'}$(LO)&B$_{2u,3'}$(LO)\\
\hline
&$i\,$175&$i\,$72&157&362&487&358&439&753\\
\cline{4-9}
&90&334& B$_{3u,1}$(TO)&B$_{3u,2}$(TO)&B$_{3u,3}$(TO)&B$_{3u,1'}$(LO)&B$_{3u,2'}$(LO)&B$_{3u,3'}$(LO)\\
\cline{4-9}
&336&725&$i\,$227&327&454&325&385&684\\
\hline
\hline
\end{tabular}
\label{tab:moden}
\end{table*}

\begin{figure*}
\includegraphics[width=0.8\textwidth]{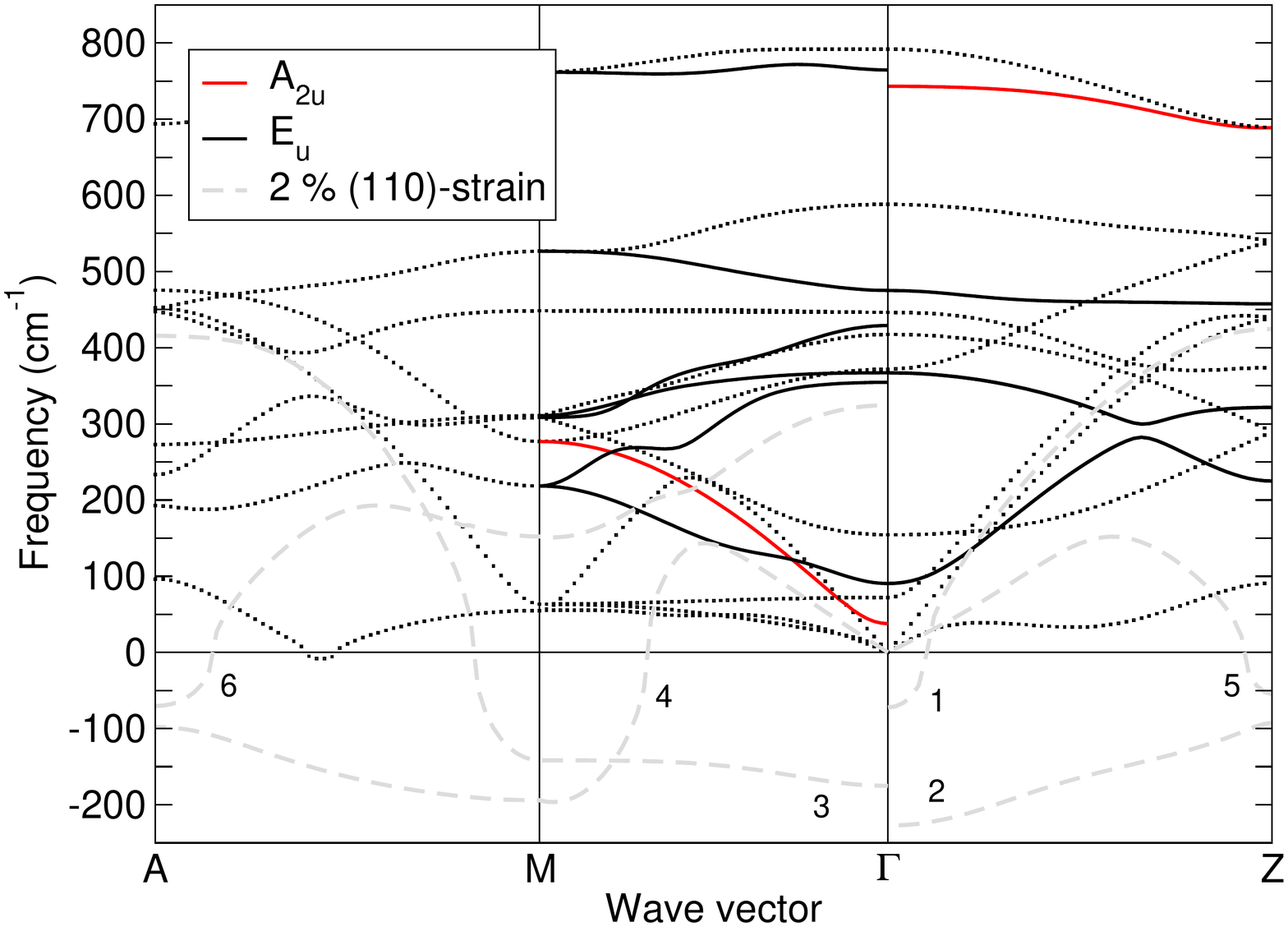} 
\caption{Phonon spectrum of TiO$_2$ rutile at the equilibrium lattice parameters obtained with PWscf (dotted lines) with highlighted 
polar modes at $\Gamma$ (solid lines).
Soft modes for 2~\% expansive (110) strain are included (dashed lines).
(1) B$_{1u}$ (LO) (2) B$_{3u}$ (TO) (3) B$_{1u}$ (TO) (4)-(5) acoustic (6) B$_{3u}$ (LO) at $\Gamma$.}
\label{fig:Phonon}
\end{figure*}

\begin{figure*}
\subfigure{
\includegraphics[width=0.75\textwidth]{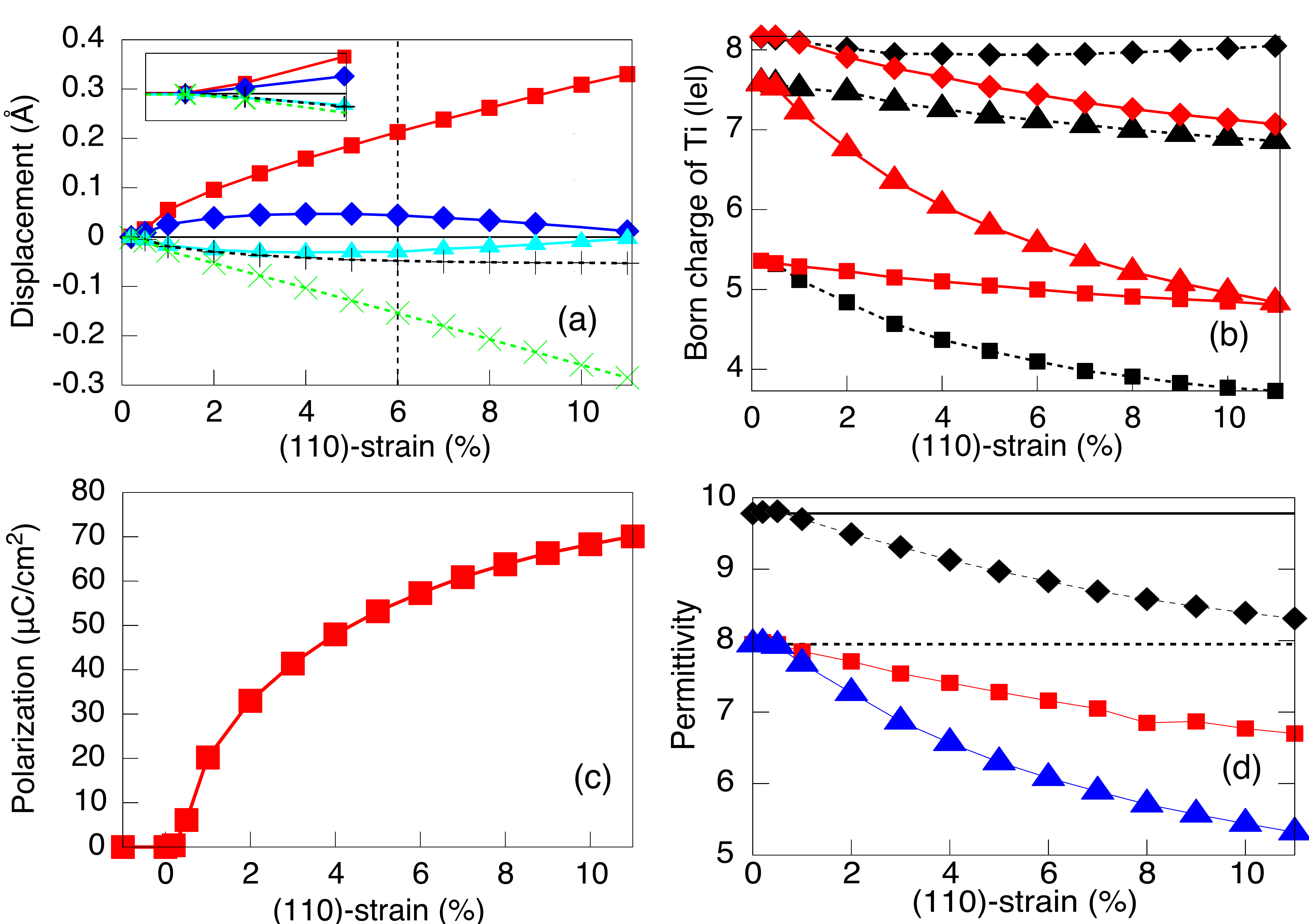} }
\caption{(a) Atomic shift of the atomic subclasses in (110) direction under expansive (110) strain. Red squares: Ti$_1$; Blue diamods: Ti$_2$; Cyan triangles: O$_1$; Black plus: O$_{2,1}$; Green crosses: O$_{2,2}$; The inset in (a) shows 0-1\% strain. (b) Born charges of Ti within the ferroelectric state for different (110) strains. Ti$_1$: Red solid curve, Ti$_2$: Black dotted curve;  Diamonds Z$^*_{(001)}$; Triangles: Z$^*_{||}$; Squares: Z$^*_{\perp}$. (c) Spontaneous polarization for different values of strain. (d) Evolution of the electronic permittivity $\epsilon_{\infty}$ within the ferroelectric state with strain. Black diamonds: (001) direction; Red squares: (110) direction; Blue triangles: ($\bar{1}10$) direction.}
\label{fig:shift}
\end{figure*}

Previous work confirmed that the ferroelectric transitions of rutile is linked to the softening of polar phonon modes.\cite{Montanari,Liu,Samara}
Because of this, the phonon spectrum of the material will be discussed in the following.
The irreducible representations of the optical vibrations of rutile at $\Gamma$ are given by symmetry\cite{Lee} as $$\Gamma_{opt}=A_{1g}+A_{2g}+A_{2u}+2B_{1u}+B_{1g}+B_{2g}+E_g+3E_u.$$
Besides the lowest frequency polar mode, $A_{2u}$, the system possesses 6 modes of E$_u$ symmetry which are polarized in the $(110)/(\bar{1}10)$ plane.
Since all these modes are doubly degenerated, the eigenenergies E$_{u,1}$, E$_{u,2}$ and E$_{u,3}$ correspond to a basis of two eigenvectors each and thus atomic displacements in the whole plane are not distinguishable.
One possible basis for the E$_{u,i}$ modes can be constructed from the eigenvectors given in Tab.~\ref{tab:ev} combined with the same displacement pattern in ($\bar{1}10$) direction. 

The lowest mode, E$_{u,1}$, consists mainly of a rigid shift between Ti and O sublattices.
Within the E$_{u,2}$ mode the Ti-sublattices are shifted against each other hereby each Ti-atom is shifted into the same direction as its equatorial O neighbors.
The highest frequency mode, E$_{u,3}$, shows exactly the opposite trend with a relative shift between the Ti sublattices and the Ti-O$_{eq}$ neighbors, while the Ti atoms and their apical neighbors are shifted into the same direction.

The double degeneracy of the E$_u$ modes at $\Gamma$ is lifted because the large dipolar interaction induces an energy penalty for the longitudinal optical modes (LO) in comparison to the transversal optical modes (TO), see Tab.~\ref{tab:moden}.\\
Besides the energetic LO-TO splitting, the dipolar interaction induces a finite coupling between the E$_u$ modes and thus modified eigenvectors appear for the LO modes, see Tab.~\ref{tab:ev}.
An alike mixing of the  TO E$_u$ modes due to the dipolar interaction has been discussed in Ref.~\onlinecite{Lee}.


Figure~\ref{fig:Phonon} shows the calculated phonon spectrum in parts of the first Brillouin zone. 
The eigenvalues obtained at $\Gamma$ are compared to literature in Table~\ref{tab:moden}.
Our calculated phonon spectrum matches qualitatively previous theoretical studies and experimental results.
As in Ref.~\onlinecite{Mitev}, we find an anomalously soft acoustic phonon mode in the Z-A-M region of the Brillouin zone.
However, it has to be noted that the absolute value of the corresponding frequency at this point is in the order of the accuracy of our calculation.
Besides this, there exist large quantitative deviations.
As the phonon frequencies are very sensitive with respect to small changes of the lattice constants, especially the lowest polar modes,\cite{Liu} these deviations can in part be explained by the slightly larger atomic distances obtained within our LDA calculations in comparison to former results.
Most notably, it has been shown\cite{Liu} that the B$_{1g}$ frequency increases with lattice expansion while the LO E$_u$ modes remain nearly unchanged.
Thus, our deviations of the mode frequencies are showing the expected trend for a small lattice expansion.

Additionally, differences in the used basis sets, the imposed sum rules and the accuracies of the simulation have drastic effects on the lowest modes. Detailed discussions for TiO$_2$ and the similar {\it{incipient}} ferroelectric material SrTiO$_3$ can, for example, be found in Refs.~\onlinecite{Shojaee, Evarestov}.
The large quantitative deviations with respect to experiments can be partially traced back to the measurement at finite temperatures since it has been shown in Ref.~\onlinecite{Traylor} that the lowest polar mode softens substantially with decreasing temperature.
 

\begin{table}
\caption{Displacement pattern of the mode eigenvectors (arbitrary units). For each atomic class, see Fig.~\ref{fig:rutil}, the amplitudes of the atomic shift in the given directions are listed. Relative amplitudes of the B$_{u,i}$ modes are given for 2~\% expansive strain.}
\begin{minipage}[b]{0.45\textwidth}
\begin{tabular}{lccccccccc}
\hline
\hline
&Ti$_1$&Ti$_2$&O$_1$&O$_2$\\
\hline
E$_{u,1}$/B$_{2u,1}$\footnote{Displacement in ($\bar{1}10$) direction}&0.1&0.3&-0.3&-0.3\\
E$_{u,2}$/B$_{2u,2}$&0.3&-0.2&0.2&-0.4\\
E$_{u,3}$/B$_{2u,3}$&0.2&-0.3&-0.2&0.4\\
\hline
E$_{u,1'}$/B$_{2u,1'}$(LO)&0.2&0.0/-0.1&0.2&-0.4\\
E$_{u,2'}$/B$_{2u,2'}$(LO)&0.3&-0.5&-0.1&0.3\\
E$_{u,3'}$/B$_{2u,3'}$ (LO)&0.2&0.2&-0.5&-0.1\\
\hline
\hline
B$_{3u,1}$(TO)\footnote{Displacement in ($110$) direction}&0.3&0.1&-0.2&-0.4\\
B$_{3u,2}$(TO)&0.1&-0.3&0.4&-0.2\\
B$_{3u,3}$(TO)&0.2&-0.3&-0.2&0.4\\
\hline
B$_{3u,1'}$LO&0.2&-0.3&0.4&-0.2\\
B$_{3u,2'}$LO&0.4&-0.2&-0.4&0.1\\
B$_{3u,3'}$LO&0.1&0.2&-0.1&-0.5\\
\hline
\hline
\end{tabular}
\label{tab:ev}
\end{minipage}
\begin{minipage}[b]{0.45\textwidth}
\begin{tabular}{lccccccccc}
\hline
\hline
&Ti$_1$&Ti$_2$&O$_1$&O$_2$\\
\hline

\hline
A$_{2u}$(LO)/(TO)\footnote{Displacement in ($001$) direction}&0.3&0.3&0.5&0.5\\
\hline
B$_{1u,1}$(TO)&0&0.6&-0.5&-0.3\\
B$_{1u,2}$(TO)&0.7&-0.3&-0.3&-0.4\\
B$_{1u,3}$(TO)&0&0.1&0.5&-0.5\\
\hline
B$_{1u,1}$(LO)&0.7&-0.7&0.1&-0.1\\
B$_{1u,2}$(LO)&0.3&0.3&-0.4&-0.5\\
B$_{1u,3}$(LO)&-0.1&0.0&0.5&-0.5\\
\hline
\hline
\end{tabular}

\end{minipage}
\label{tab:ev}
\end{table}


As discussed in the introduction, the softening of the polar A$_{2u}$ mode under lattice expansion or expansive (001) strain induces a ferroelectric transition.
This effect can be explained by simple energy arguments.
The relative stability of the ferroelectric and paraelectric phases depends on the dipolar energy gained by the ferroelectric state and the energy penalty due to the short range repulsion which builds up for the relative atomic displacement.
As the equatorial Ti-O bonds contribute most to the short-range repulsion, a ferroelectric transition becomes likely if this distance is sufficiently enlarged.
Similarly, a uniform lattice expansion causes a softening of the E$_u$ modes,\cite{Montanari,Liu} although this effect is less pronounced than for the (001) direction.
Additionally, one may expect softening of the polar modes in (110) and (001) direction for expansive (110) strain in this picture.

\subsection{(110) strain}
Besides the modification of the Ti-O distances, strain in (110) direction reduces the symmetry of the system to the orthorhombic $C{mmm}$ class.
Two inequivalent Ti and O sublattices form, see Fig.~\ref{fig:rutil}(b). Within the first subclass the short Ti-O distances are conserved whereas the equatorial bond length within the second subclass is modified. The optical modes at $\Gamma$ are now represented by\cite{Bilbao}
$$\Gamma_{opt}=2A_{g}+2B_{1g}+B_{2g}+B_{3g}+3B_{1u}+3B_{2u}+3B_{3u}$$
and the degeneracy of the E$_u$ modes is lifted to B$_{2u,i}$ in ($\bar{1}10$) direction and  B$_{3u,i}$ in the strained (110) direction. 
An alike mode splitting appears for the acoustic modes and thus the destabilization of the atomic structure along the acoustic mode in (110) direction has been predicted.~\cite{Mitev}\\
Although, we find an alike softening of acoustic modes (branches (4)-(5) in Fig.~\ref{fig:Phonon}), this phonon branches show a large mode mixing and are both polar in (110) direction within the soft regions. Additionally, our previous investigations with VASP yield a destabilization of the system towards a ferroelectric state under expansive strain caused by the softening of a polar mode.\cite{110}\\
In the following, the atomic displacements and the resulting ferroelectric state will be discussed in more detail.
Within these simulations no diagonalization of the dynamic matrix has been performed. Instead the paraelectric symmetry of the system has been broken by a shift of one O-atom within a $\sqrt{2}\times\sqrt{2}\times1$ supercell, see Fig.~\ref{fig:rutil}. The atomic positions have been optimized within this 
 $Pmm2$ symmetry class and thus, the O$_2$ atoms are split into two inequivalent subclasses.\\
 The amplitude of the obtained polar shift increases continuously with imposed expansive strain, see Fig.~\ref{fig:shift}(a).
 Already at 0.5~\% strain, the amplitude of the displacement along the polar mode is exceeding the arbitrary displacements below the given accuracy of the atomic relaxation.
Between 2~\% and 6~\% strain, the displacements of Ti$_1$ and O$_{2,2}$ atoms increase approximately linearly, while the shift of the Ti$_2$ and O$_1$ atom is nearly constant within this range of strain and the increase in the  O$_{2,1}$ displacement is only minor.
For larger values of strain the Ti$_1$ amplitudes start to saturate, while the O$_1$ and Ti$_2$ displacements even decrease.
This evolution of the atomic displacements with imposed strain is a fingerprint of a second order phase transition. Although, the exact transition strain cannot be determined, the transition seems to start for infinitesimal expanding strain.
Most notably, the spontaneous polarization shows the characteristic profile of such a phase transition with a continuous onset and a saturation for large values of strain, see Fig.~\ref{fig:shift}(c).

The saturation of the spontaneous polarization can again be explained by simple energy arguments.
While the reduction of the short range repulsion further increases with large values of strain, the dipolar energy is reduced in this case.
As Born effective charges are proportional to the dipolar interaction, this argument can be confirmed by the evolution of the Born charges in (110) direction, Z$^*_{(110)}$, under strain, see Fig.~\ref{fig:shift}(b).  
The tensor of the Born charges is diagonal within a $(\bar{1}10)$, $(010)$, $(001)$ reference system, with the principal values Z$^*_{(001)}$ aligned along (001) direction,  Z$^*_{||}$ aligned along the apical bond and Z$^*_{\perp}$ perpendicular to both other directions. For a detailed analysis, see Refs.~\onlinecite{felich,Cangiani}.

For Ti$_1$ atoms, the apical Ti$_1$-O$_2$ bonds are aligned along (110) and thus the large dipolar interaction leads to a large shift of the corresponding Ti-atoms, cf.  Z$^*_{||}$. 
Instead, the apical Ti$_2$-O$_1$ bonds are aligned along ($\bar{1}$10) and thus the dipolar interaction in (110) is much smaller, cf. Z$^*_{\perp}$.
For both Ti classes, the Born charge along (110) decreases considerably with increasing strain as does the dipolar interaction.
Especially, for the Ti$_2$ atoms the Born charge is reduced to the nominal charge of Ti$^{4+}$ at about 6~\% strain and thus vanishing dipolar interaction and atomic shifts appear for increasing strain.

In order to characterize further the properties of the ferroelectric state, its electronic permittivity, $\epsilon_{\infty}$, for different values of strain is shown, see Fig.~\ref{fig:shift}(b).
Although, a large permittivity is obtained over the whole investigated interval starting form 0~\% to 11~\% expansive strain, all principle values of the permittivity tensor decrease more than linearly with increasing strain.
Especially, $\epsilon_{\infty}$ in ($\bar{1}10$) is considerably reduced.\\
Additionally, the influence of the strain induced ferroelectric state on the electronic structure has been investigated.
For this purpose, the density of states has been analyzed for the ferroelectric state at 5~\% strain.
As only small deviations in comparison to the undistorted material appear,  similar electronic properties have to be expected. Only, the obtained band gap is slightly enlarged to 1.82~eV in comparison to the band gap of 1.64~eV for the undistorted bulk.


If the polar shift is prevented by the imposed $Cmmm$ symmetry, or due to the large short range repulsion which appears under compressive strain, only the oxygen atoms relax and thus the internal parameter $u$ is modified. This relaxation corresponds to a change of the relative Ti-O$_{eq}$-Ti-O$_{ap}$ distances due to the larger compressibility of the apical bond in contrast to the equatorial bond.\cite{Montanari} 
Hence, the apical bond length is further enlarged for expansive strain, while the strain-imposed increase of the equatorial bond length is reduced (vice versa for compressive strain).
Qualitatively the same relaxations are obtained within VASP and PWscf simulations.
The main relaxation can be linked to a A$_{1g}$ mode of the undistorted system for which we yield a frequency of 588 cm $^{-1}$ at $\Gamma$. 
Since the strain induced modification of the Ti-O distances is not uniform in (110) direction and ($\bar1$10), the relaxation of the internal parameter differs in both directions and thus a modification of the relaxation amplitude for the O-subclasses is superimposed.

In Ref.~\onlinecite{110} we have shown that the calculated ferroelectric displacement pattern can be approximated by a simplified mixture of the bulk E$_u$ modes with unique amplitudes.
In agreement with this finding, the diagonalization of the dynamic matrix at 2~\% expansive strain, based on PWscf, yields an imaginary phonon frequency for the lowest B$_{3u}$ mode.
The eigenvector, see Tab.~\ref{tab:ev}, of this mode is qualitatively matching our earlier results obtained with VASP, as the main shift consists of a relative shift between the Ti and O atoms with superimposed relative shifts between the Ti and O atoms corresponding to different subclasses.
Note, that the PWscf results have been obtained for the primitive cell wit $Cmmm$ symmetry in order to reduce the computational time and thus the O$_2$ atoms are all symmetrically equivalent.
Therefore, the appearance of the discussed ferroelectric state with (110) polarization can be qualitatively reproduced for different simulation details.

Besides the large softening of the lowest B$_{3u}$ mode, all other B$_{3u}$ modes are shifted to lower energies at $\Gamma$ due to the discussed lifting of the degeneracy.
In contrast, the B$_{2u}$ modes are shifted to higher frequencies.
The only exceptions are the  B$_{2u,2}$ and B$_{2u,3'}$ modes which are also lowered in comparison to the undistorted bulk.

The strain imposed modification of the eigenenergies is accompanied by a modification of the mode eigenvectors for the $B_{3u}$ modes, while the $B_{2u}$ eigenvectors are identically to the corresponding E$_u$ modes of the undistorted system within the given accuracy. 

The relative amplitudes within the polar modes can again be qualitatively understood by the counterplay of dipolar interaction and short range repulsion.
As previously discussed for the (110) direction, the Ti atoms belonging to class 1 contribute more to the dipolar interaction than the Ti$_2$ atoms. The opposite trend holds for the shift in ($\bar{1}$10) direction, where a larger contribution to the dipolar interaction is obtained for the second subclass.
Additionally, the equatorial Ti$_2$-O$_2$ distance is enlarged due to the imposed strain and thus the short range repulsion is reduced for a relative shift between the Ti$_2$ and O$_2$ atoms.
Both effects lead to a larger  Ti$_2$-O$_2$  shift in comparison to the Ti$_1$-O$_1$ shift for the B$_{2u}$  modes.
Although, a similar short-range repulsion is obtained for the B$_{3u}$ modes, the different dipolar interaction leads to modified relative amplitudes.\\
Besides, the discussed softening of the B$_{3u}$ modes, our previous investigation yields a softening of the A$_{2u}$ mode under expansive (110) strain,\cite{110} if the polar displacements in (110) direction are prevented by the imposed symmetry.
In contrast to this, the mode has been predicted to be quiet insensitive towards (110) strain.\cite{Mitev}\\
For the strained system the single polar mode in (001) direction, the A$_{2u}$ mode, is replaced by 3 polar B$_{1u}$ modes.
The corresponding mode eigenvectors are listed in the lower part of Tab.~\ref{tab:ev}.  While the A$_{2u}$ LO and TO modes are characterized by a shift of the Ti-O sublattices with a unique amplitude, relative shifts between the Ti- and O subclasses are superimposed for the B$_{1u}$(TO) modes.
In Ref.~\onlinecite{110} we assumed unique amplitudes for the atomic displacements and thus the atomic shift along the real mode pattern is most likely even more favorable.\\
The diagonalization of the dynamical matrix for 2~\% strain yields imaginary frequencies for the (LO) and (TO) B$_{1u,1}$ modes. Therefore, our results obtained with the PWscf code confirm our previous findings that the ferroelectric state polarized in (001) direction is energetically more favorable than the paraelectric state although less favorable than the ferroelectric state polarized in (110) direction.\cite{110}\\
It is most likely that a ferroelectric transitions along (001) takes place along the B$_{1u}$ (TO) eigenvector.
For this eigenvector, the relative shift between the Ti$_2$-O$_2$ equatorial bonds is enlarged in comparison to the shift of the  Ti$_1$-O$_1$ neighbors, due to the different short range repulsion.
 If the evolution of the  dipolar interaction is again monitored by the Born charges, an even larger modulation of the atomic shifts has to be expected for increasing strain, as Z$^*_{(001)}$ is decreasing for Ti$_1$  with increasing strain, while the dipolar interaction is nearly constant for the Ti$_2$ atoms.\\
As in SrTiO$_3$ \cite{Antons} a large coupling between the soft modes in (110) and (001) directions exists and the atomic displacement along the B$_{1u}$ modes is disabled within the polar state in (110) direction.
This means, the phonon frequencies for an polar displacement along (001) are no longer imaginary if the atoms are displaced along the B$_{3u}$ eigenvector.

\section{Conclusions and Outlook}
In this paper we have discussed the softening of polar modes  under (110) strain in rutile, finding a second
order phase transition (polarization versus (110) strain) for a polar state in (110) direction. The calculations were done using the 
plane-wave method as implemented in the VASP and PWscf codes.
The two methods allowed for a cross-checking of the results and
lead to qualitative agreement regarding the atomic relaxations and atomic mode patterns
under (110) strain.
The decomposition of the displacement pattern "under strain" into ideal 
$E_u$ modes is qualitatively matching  the eigenvectors of the
lowest $B_{3u}$ mode and a detailed analysis of the polar modes and 
associated eigenvectors has been undertaken. The investigation
of displacements, dipolar interaction, dielectric permittivity
and electronic structure underlines the trend for a second order
ferroelectric transition under (110) strain in rutile TiO$_2$.
The calculations have been complemented by evaluating the
phonon dispersion curves of unstrained rutile and the appearance
of imaginary branches 
at 2~\% strain, which correspond to polar displacements in (001) and (110) direction.
The coupling of these polar $B_{1u}$ and $B_{3u}$ modes as well
as their different dependencies on (001) and (110) strains
leads to the possibility of strain engineering. Thus, a phase
diagram for the relative stability of both kinds of modes at
different combinations of strains has to be calculated in future.
In addition, it will be necessary to calculate the phonon spectra of 
both ferroelectric phases and to characterize their interrelation.
This should be supplemented by evaluating the electron-phonon
coupling constant and its impact on the structural transitions,
which may occur in strained rutile. Finally the relation to
experimental setups has to be achieved.
Therefore, calculations on strained surfaces will be performed in order to investigate the influence of the surface and surface induced atomic relaxations on the polar modes in thin films of TiO$_2$.        

\section{Acknowledgments}
This work was supported by the Deutsche Forschungsgemeinschaft (SFB 445, SPP 1239) and by Science Foundation Ireland (SFI-07/YI2/I1051).
Computational time has been provided by the John von Neumann Institute for Computing.

\bibliography{../Fe_rutil110/anna}

\end{document}